\documentclass[12pt]{article}
\usepackage{graphicx,amssymb}
\usepackage{pazh}
\tightenlines

\sloppypar

\begin{document}

\title{\bf Gravitational Wave Bursts from Collisions 
of Primordial Black Holes in Clusters}

\author{\bf \hspace{-1.3cm} \ \
V.I. Dokuchaev\affilmark{1}, Yu.N. Eroshenko\affilmark{1},
S.G. Rubin\affilmark{2}}

\affil{
{\it Institute for Nuclear Research RAS, 
pr. 60-letiya Oktyabrya 7a, Moscow, 117312 Russia}$^1$\\
{\it Moscow Engineering Physics Institute, Kashirskoe sh. 21, Moscow, 115409 Russia }$^2$}

\sloppypar \vspace{2mm} \noindent The rate of gravitational wave bursts from the mergers of massive primordial black holes in 
clusters is calculated. Such clusters of black holes can be formed through phase transitions in the early 
Universe. The central black holes in clusters can serve as the seeds of supermassive black holes in 
galactic nuclei. The expected burst detection rate by the LISA gravitational wave detector is estimated. 



\noindent {\bf DOI:\/} 10.1134/S1063773709030013 

\section*{INTRODUCTION}

The Laser Interferometer Space Antenna (LISA) 
designed to detect gravitational waves is scheduled to 
be launched in 2015 (for details, see the official LISA 
site). In comparison with ground-based detectors, 
LISA will be able to record signals with relatively 
low frequencies, $10^{-4}-1$~Hz. The mergers of massive 
black holes (BHs) in galaxies can be the sources of 
such events. This process is very likely, since many of 
the suggested models for the formation of supermassive
BHs include multiple mergers of intermediatemass
BHs as a necessary stage (Cattaneo et al. 
2005; Volonteri et al. 2007; Di Matteo et al. 2007). 
The suggested scenarios predict different BH distributions
and, accordingly, different BH merger rates. 
It is hoped that future observations with LISA will 
help narrow the range of possible models and identify 
the most likely ones by characteristic features in the 
distribution of gravitational wave bursts. 

The origin of supermassive BHs is an especially 
challenging problem. In particular, the existence 
of supermassive BHs in quasars at high redshifts, 
$z\simeq6-7$, requires an explanation. These supermassive
BHs appeared very early and, hence, there existed 
an efficient mechanism of their birth and subsequent 
rapid growth in a relatively short cosmological time 
interval. Several models for the formation of supermassive
BHs have been suggested (for a review of 
the various astrophysical and cosmological scenarios, 
see, e.g., Dokuchaev et al. 2007). One of the most interesting
research directions is based on cosmological 
formation mechanisms of massive primordial BHs 
at very early pregalactic epochs, which can explain 
the existence of BHs at high redshifts following from 
observations. 

Our calculations are based on the formation 
mechanism of massive primordial BHs at very early 
radiation-dominated epochs. This mechanism was 
suggested and described in detail by Rubin et al. 
(2000, 2001), Khlopov and Rubin (2004), and Khlopov
et al. (2005). Briefly, its essence is as follows. In 
inflationary models using potentials with more than 
one minimum, second-order phase transitions occur 
even at the inflationary stage. Spatial regions that 
will be at a minimum of the potential at the end of 
inflation different from the minimum to which the 
surrounding space evolves are formed. In contrast, 
at the inflationary stage, the size of these regions 
increases exponentially. The fundamental point is that 
after the completion of inflation, the exponentially 
expanded region has a size much larger than that of 
the horizon size (at that time). In the above papers, 
these regions were also shown to be surrounded by a 
closed field wall with some surface energy density. 
The energy contained in the wall is proportional 
to the square of its size and can vary over a wide 
range, since its size is not limited by the horizon size. 
After crossing the horizon to the post-inflationary 
stage, the wall evolves rapidly and collapses into a 
primordial black hole (PBH). The PBH mass depends 
on the parameters of the inflaton potential and initial 
conditions. We chose the conditions under which the 
number and mass of PBHs were consistent with the 
observational data. More specifically, the mass of the 
most massive PBHs formed via the collapse of the 
field wall should be $10^5M_{\odot}$ 
and the energy density in the PBH does not exceed 1\% of the dark energy 
density. Rubin et al. (2001) showed that the mass 
spectrum of the most massive PBHs falls off rapidly 
with mass and we chose $10^5M_{\odot}$
as a typical value, which is sufficient for the subsequent estimations. 

Based on the formalism of these papers, we constructed
a model for the formation of protogalaxies 
(from dwarf to supermassive ones) induced by massive
PBHs (Dokuchaev et al. 2005, 2008). This paper, 
in which we consider the gravitational wave bursts 
from the mergers of PBHs in clusters, is directly 
based on the results of our previous works on the 
formation of BH clusters (Dokuchaev et al. 2005, 
2008). The protogalaxy formed around a BH cluster 
turns out to have a number of realistic properties that 
correspond to the observed properties of galaxies: the 
presence of a central supermassive BH, the existence 
of intermediate-mass BHs in an extended halo, and 
asufficient amount of dark and baryonic matter to 
produce the observed number of stars in galaxies. An 
important characteristic feature of the suggested scenario
is a very early formation of quasars (Dokuchaev 
et al. 2005). 

LISA provides a unique opportunity to test the 
scenarios for the formation of supermassive BHs 
in galactic nuclei. The possibility of such testing is 
related to the inevitable BH mergers that generate 
bursts of gravitational radiation. The merger rate 
of BHs in the cosmological PBH formation model 
under consideration differs significantly from that of 
BHs formed through astrophysical mechanisms. As a 
typical case, we consider a cluster that we calculated 
based on the papers of Khlopov et al. (2005) and 
Dokuchaev et al. (2005, 2008). There is a supermassive
BH with a mass $M_0\sim10^5M_{\odot}$ surrounded by 
lower-mass BHs at the center of such a characteristic 
cluster. The distributions of intermediate-mass BHs 
obtained from this formalism (Klopov et al. 2005) will 
be used below as the initial data. For convenience, 
we found a fit to the BH mass distribution. More 
specifically, after the virialization of the PBH cluster 
at the radiation-dominated epoch, the differential BH 
number density in the cluster can be represented as 
\begin{equation}
\frac{dn}{dM}= 1.6\times10^{3}
\left(\frac{r}{1\mbox{~pc}}\right)^{-3}\left(\frac{M}{M_{\odot}}
\right)^{-2}\mbox{$M_{\odot}^{-1}$pc$^{-3}$}. \label{bhfit1}
\end{equation}
Another dynamical component of the clusters that 
should be taken into account when the structure of
protogalaxies is investigated is dark matter. The total 
mass of the dark matter is higher than that of the 
PBHs at distances larger than $r\simeq1.6$~pc from the 
cluster center. However, in this paper, we will restrict 
our analysis only to the central region of the cluster 
where the BH mass dominates over the dark matter 
mass. 

\section*{COLLISIONS OF BLACK HOLES 
IN CLUSTERS}

Dynamical relaxation takes place in the densest 
central region of the cluster under the effect of twobody
gravitational BH scatterings. As a result of this 
relaxation, the intermediate-mass BHs surrounding 
the most massive central BH occasionally fall into 
the loss cone and merge with the central BH. The 
intermediate stage of these mergers is the formation 
of short-lived close BH pairs. The cross section for 
the gravitational capture of two BHs with masses $M_0$
and $M$ into a close pair (Mouri and Taniguchi 2002) is 
\begin{equation}
\sigma_{\rm mer}=2\pi\left(\frac{85\pi}{6\sqrt{2}}\right)^{2/7}
\frac{G^2(M_0+M)^{10/7}M_0^{2/7}M^{2/7}} {c^{10/7}v_{\rm
rel}^{18/7}}, \label{m1m2}
\end{equation}
where $v_{\rm rel}$ is the relative BH velocity; $G$ and $c$ are the 
gravitational constant and the speed of light, respectively.
After their capture, the BHs merge together 
relatively fast through the radiation of gravitational 
waves. A powerful gravitational wave burst is generated
at the time of the merger. The cross section 
for the merger of two BHs with the formation of an 
intermediate close pair in a wide range of BH masses 
and velocities exceeds significantly the cross section 
for direct BH collisions with gravitational focusing. 

Let us analyze the main parameters of the collisions
 between a BH with a mass $M$ and a central 
BH at rest with a mass $M_0>M$ in the cluster under 
consideration. The differential BH merger rate is 
\begin{equation}
\label{ddN} d\dot N=\sigma_{\rm mer}v_{\rm rel}\,dn,
\end{equation}
where the number density $dn$ of BHs with masses in 
the interval $M$-$M+dM$ depends on the distance 
to the cluster center and the time. The pair relaxation 
through distant Coulomb scatterings distorts the initial
spatial distribution (\ref{bhfit1}). The pair relaxation time 
for BHs with mass M in a multicomponent system is 
(Spitzer and Saslaw 1966) 
\begin{equation}
t_{\rm rel}\simeq\frac{1}{4\pi}\frac{v^{3}}{G^{2}M
\Lambda_c\rho_h}, \label{trel}
\end{equation}
where $\rho_h$ is the total density of the BHs of all masses, 
$v$ is the velocity dispersion in the cluster, and $\Lambda_c\simeq15$ is the Coulomb logarithm. The relaxed spherical subsystem
of BHs persists in the cluster for $\sim40t_{\rm rel}$ until 
its dynamical evaporation. Some of the BHs merge 
with the central most massive BH. The subsystems of 
an increasingly large scale with a lower density relax 
in the course of time. Consequently, the BH number 
density in the central region, where the most massive 
BH is located, falls and the merger rate gradually 
decreases. During our calculations, we first find the 
size of the completely relaxed cluster region for each 
cosmological time $t$, where $t\simeq40t_{\rm rel}$. We use the 
average BH number density in the relaxed region (the 
initial number density (1) is averaged) as the number 
density dn in Eq.~(3) and assume for our estimation
that the density in this region is uniform. We 
perform our calculation with a logarithmic accuracy 
by neglecting the change in the Coulomb logarithms 
compared to the power-law dependences. In this approximation,
we obtain the following expression for 
the BH merger rate in one cluster: 
\begin{equation}
\frac{d\dot N(z)}{dM}\simeq5.1\times10^{-7}\left(\frac{M}{M_{\odot}}
\right)^{-67/21}\left(t(z)/t_0\right)^{-31/21}M_{\odot}^{-1}\mbox{~yr}^{-1},
\label{gfit}
\end{equation}
where $t(z)$ is the redshift dependence of the cosmological
time calculated from well-known formulas and 
$t_0$ is the present time. In reality, mass segregation 
takes place in the cluster: high-mass BHs settle toward
the cluster center and their merger rate will be 
slightly higher than the calculated one. Therefore, our 
calculations should be considered as a lower limit. 

A gravitational detector records bursts based on 
an optimal filtering technique with an efficiency dependent
on the signal-to-noise ratio (Grishchuk et al. 
2001). Let us find the signal-to-noise ratio $\rho_{\rm SN}$ for 
the detection of BH mergers in the clusters under 
consideration by LISA. Our approach is similar to 
that used by Will (2004), but we apply cosmological 
corrections, since the redshift can be much higher 
than unity. The signal-to-noise ratio as a function of 
the BH mass and the cluster redshift is 
\begin{equation}
 \rho_{\rm SN}(z,M)^2=
 4\int\limits_{f_i}^{f_f}\frac{|\tilde h(f)|^2\,df}{S_h(f)},
 \label{sn}
\end{equation}
where $f_i$ and $f_f$ are, respectively, the initial (minimum)
and final (maximum) frequencies of the observed
gravitational radiation, $\tilde
h(f)$ is the Fourier 
transform of the observed signal, and $S_h(f)$ is the 
detector noise spectral density. 

Denote the sum of the merging BH masses by 
$\tilde M=M_0+M$. The orbit of the BH pair decays due to 
energy losses through gravitational radiation and the 
orbital revolution frequency increases. The minimum frequency of the gravitational signal corresponds to 
the initial time of observations, while the maximum 
frequency is twice the revolution frequency in the 
innermost stable orbit $f_f=f_m/(1+z)$, where (Will 
2004) 
\begin{equation}
 f_m\simeq\frac{c^3}{6^{3/2}\pi G\tilde M}.
 \label{ff}
\end{equation}
For example, $f_m=0.04$~Hz for $\tilde M=10^5M_{\odot}$.
Over the observing time $T=1$~yr (and, accordingly, over 
the proper time of the source $T/(1+z)$), the orbital 
revolution frequency and the gravitational radiation 
frequency increased from the initial value of (Will 
2004; without the factor $1+z$) 
\begin{equation}
f_i=f_f\left(1+\frac{256\pi^{8/3}}{5} \frac{G^{5/3}M_0M}{c^5\tilde
M^{1/3}}f_m^{8/3}\frac{T}{1+z}\right)^{-3/8}. \label{uorb}
\end{equation}
The observed gravitational wave spectrum is well fitted
 by a power law $\tilde h(f)=Af^{-7/6}$ for $f<f_f$ and 
$\tilde
h(f)=0$ for $f>f_f$ (Will 2004), where 
\begin{equation}
A=\frac{1}{\sqrt{30}\pi^{2/3}}
\frac{G^{5/6}(M_0M)^{1/2}(1+z)^{1/3}}{d_Lc^{3/2}\tilde M^{1/6}}.
\end{equation}
Here, $d_L$ is the luminosity distance. We will use an 
analytical fit to the noise spectrum from Finn and 
Thorne (2000): 
\begin{equation}
S_h(f)=6.12\;10^{-51}/f^4+1.06\;10^{-40}+6.12\;10^{-37}f^2
\mbox{~Hz}^{-1}. \label{sfit}
\end{equation}

\begin{figure}
\begin{center}
\includegraphics[angle=0,width=0.9\textwidth]{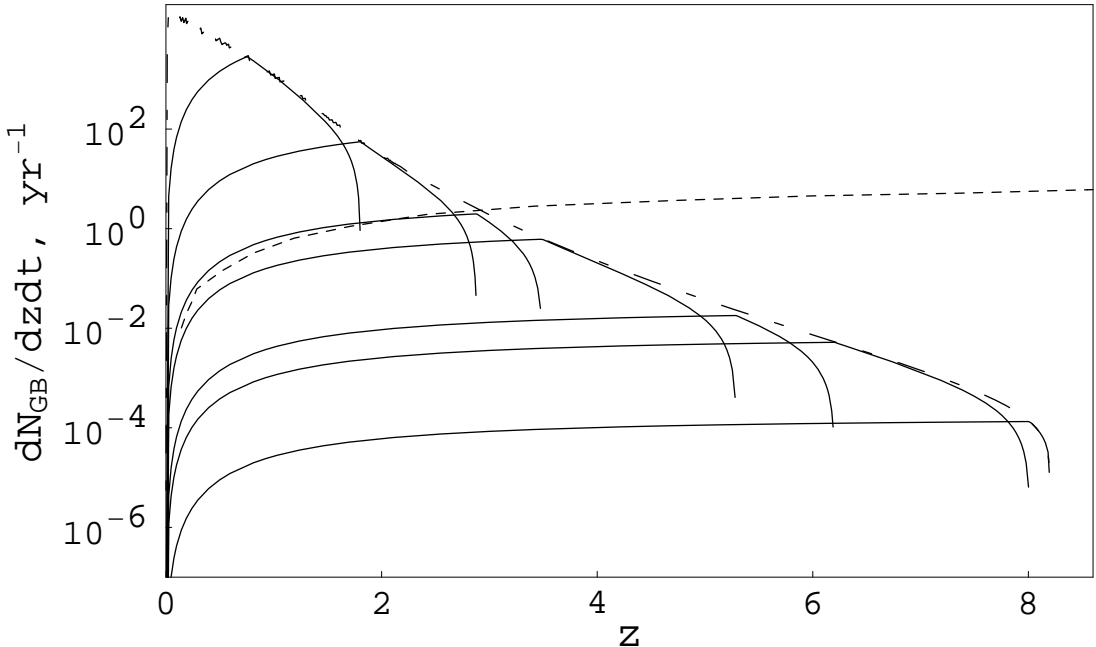}
\end{center}
\caption{Predicted distribution of gravitational wave bursts in redshift z at which the signal sources, the clusters of primordial 
BHs, are located. The solid curves correspond to the total merger rates for BHs from the mass intervals $M=10^2-5\times10^2M_{\odot}$,
$5\times10^2-10^3M_{\odot}$, $10^3-5\times10^3M_{\odot}$,
$5\times10^3-10^4M_{\odot}$, $10^4-5\times10^4M_{\odot}$ and
$5\times10^4-10^5M_{\odot}$ 
(from the top down). The value
of $z$ at the right boundaries of the curves is the maximum redshift from which the detection of bursts by LISA is possible. 
The enveloping dash–dotted curve indicates the total rate of bursts from the mergers of BHs with masses in the range 
$10^{-1}-10^5M_{\odot}$. For comparison, the dashed line indicates the results of calculations from Sesana et al. (2005) for the mergers 
of BHs formed through the collapse of gaseous clouds. } 
\end{figure}

The observed rate of bursts from redshifts $<z$ is 
\begin{equation}
d\dot N_{\rm GB}(z)=\int\limits_{0}^{z}dz(dl_c/dz)4\pi
l_c^2n_{cl}\,d\dot N(z)(1+z)^{-1}\theta(\rho_{\rm
SN}(z,M)-5),\label{rate1}
\end{equation}
where $n_{cl}$ is the cluster number density, $\theta$ is the Heaviside
function, and, as is commonly done for LISA, 
we chose $\rho_{\rm
SN}(z,M)=5$ as the threshold of reliable 
detection. For each specific mass $M$, the detection 
threshold implicitly defines the maximum redshift $z$ 
from which the signal can be received. The quantity 
$d\dot N(z)$ in Eq.~(\ref{rate1}) corresponds to some small mass 
interval $dM$ and we will present the final results for $\dot N_{\rm
GB}(z)$ integrated over finite mass intervals. In the 
above formulas, the cosmological distance in comoving
coordinates and the luminosity distance are, respectively,
\begin{equation}
l_c(z)=\int_0^zc\,dz/H(z), \quad d_L(z)=(1+z)l_c(z),
\end{equation}
where $H(z)=\dot a/a$ is the Hubble constant; $a$ and 
$\dot a$ are the scale factor of the Universe and its time 
derivative, respectively.

To determine the cluster number density $n_{cl}$ in 
a unit comoving volume, we will use the following 
estimate. Suppose that the BH clusters have settled 
toward the galactic centers by now and that their central
BHs have merged into supermassive BHs with 
masses $\sim10^8M_{\odot}$. We will then find that there were 
initially $N_{cl}\sim500$ clusters in each structured galaxy. 
Since the number density of structured galaxies 
containing supermassive BHs is $n_g\sim0.01$~Mpc$^{-3}$, 
the cluster number density is $n_{cl}\sim500\times0.01=5$~Mpc$^{-3}$.

Figure~1 shows the distribution of gravitational 
wave bursts in redshift $z$ that we derived in comparison
with the results of Sesana et al. (2005), who 
considered the mergers of intermediate-mass BHs 
formed according to a standard astrophysical scenario
(see Fig.~3 from the above paper). As we see 
from the figure, the rate of BH mergers in the lower 
mass range is high in our model, because they are numerous.
In one-year-long observations, the detection 
of several mergers of BHs with masses in the range 
$10^2-10^3M_{\odot}$ with the central BHs of the clusters is 
probable. The bends on the plot are attributable to 
the presence of a $\theta$-function in (\ref{rate1}). These bends 
correspond to the maximum redshifts from which the mergers of minimum-mass BHs from each range are 
seen. 

Note the significant difference, obvious from Fig.~1, 
in the shapes of the burst distribution in our scenario
and the distribution in Fig.~3 from Sesana 
et al. (2005). According to this paper, BHs are formed 
through dissipative collapse of the central regions 
of gaseous clouds; subsequently, the protogalaxies 
coalesce and the BHs settle toward the new dynamical
center and merger together. Obviously, the 
BH merger rate in this model increases with time, 
at least at the first stage. In our model, primordial 
BHs were formed at the radiation-dominated epoch 
of the Universe and the dynamical evolution of dense 
BH clusters began at the same epoch. Therefore, the 
BH merger rate in our model decreases with time. 
The significant merger rate in our model compared 
to astrophysical scenarios is attributable to the high 
BH number density at the cluster centers. Thus, the 
mergers in clusters and the pair mergers in the model 
by Sesana et al. (2005) occur under fundamentally 
different conditions, which leads to the difference 
between the distributions in Fig.~1. This difference 
during LISA observations can help in choosing 
between the various scenarios for the BH origin. 

To interpret Fig.~1, it is also useful to plot the 
observed dimensionless gravitational wave amplitude 
$h_c$ against the observed frequency $f$. According to 
Sesana et al. (2005), $h_c=h\sqrt{n}$ for $n<fT$ and $h_c=h\sqrt{fT}$ for
$n>fT$, where $h$ and $n$ depend on $f$, the 
distance $l_c(z)$, and the masses of the merging BHs. 
The corresponding expressions were given by Sesana 
et al. (2005; see Eqs.~(2)–(7)). Figure~2 presents the 
curves for $z=1$; for other $z$, the curves can be easily 
recalculated by taking into account the fact that $h_c\propto l_c^{-1}(z)$. 

\begin{figure}
\begin{center}
\includegraphics[angle=0,width=0.9\textwidth]{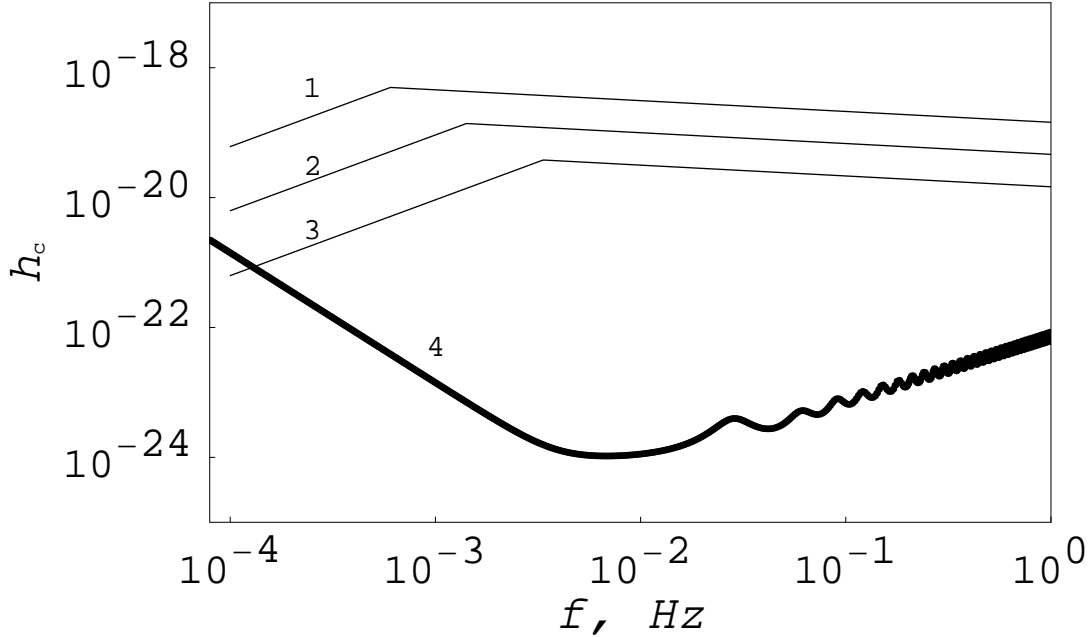}
\end{center}
\caption{Observed dimensionless gravitational wave amplitude versus frequency when the BHs in a cluster merge with a 
central BH with a mass $M_0=10^5M_{\odot}$
at redshift $z=1$. Lines 1–3 correspond to the masses of the second merging 
BH, $M_1=10^4M_{\odot}$, $10^3M_{\odot}$ and $10^2M_{\odot}$. Line 4 represents the LISA sensitivity curved as constructed from Online Curve 
Generator data (http://www.srl.caltech.edu/). } 
\end{figure}

Let us now consider the recoil impulse gained by 
the central BH as it collides with lower-mass BHs. 
This effect can lead to the BH ejection from the cluster 
if BHs of comparable masses with a high angular 
momentum collide. Damour and Gopakumar (2006) 
found the recoil velocity due to asymmetric gravitational
radiation to be at a maximum at a BH mass 
ratio $\eta\simeq0.38$ and to decrease approximately as $\eta^2$. 
Note, however, that the probability of such collisions 
in a separate cluster in our model is very low and that 
the high rate of gravitational wave bursts is related to 
the large number of clusters in the observable Universe.
Indeed, it is easy to find from Eq.~(\ref{gfit}) that the 
collision rate between BHs with masses $10^4-10^5M_{\odot}$. 
and the central BH is $\sim2\times10^{-19}$~yr$^{-1}$. The 
collision rate between the central BH and lower-mass 
BHs is higher, but the collision probability in the 
Hubble time is $\sim0.005$ even for $M\sim100M_{\odot}$ 
and the recoil velocity is approximately four orders of magnitude
lower than the velocity dispersion in the cluster. 
Thus, the recoil through gravitational radiation has 
virtually no effect on the rate of gravitational wave 
bursts in our model. Another source of the recoil 
velocity is the impulse of an impinging BH. Through 
this effect, the central BH will also gain a very low 
recoil velocity $\sim v(M/M_0)$, where $v$ is of the order 
of the velocity dispersion in the cluster $\sim50$~km~s$^{-1}$, 
and such collisions do not lead to the ejection of the 
central BH from the cluster. 

We calculated the rate of BH mergers with the 
central, most massive BH in the cluster. Let us now 
show that the contribution from these mergers is 
dominant compared to that from the mergers of BHs 
of lower masses $M_1<M_0$ and $M_2<M_0$ with one 
another. Indeed, the merger rate in a volume element
$dV$ is 
\begin{equation}
\frac{d\dot N}{dM_1dM_2}=\sigma_{\rm mer}v_{\rm
rel}\frac{dn_1}{dM_1}\frac{dn_2}{dM_2}dV. \label{rate2}
\end{equation}
Using distribution (\ref{bhfit1}) and integrating Eq. (\ref{rate2}) over 
the cluster volume and the BH masses, we will obtain 
\begin{equation}
\int\int\int 4\pi r^2\,dr\,dM_1\,dM_2\frac{d\dot
N}{dM_1dM_2}\propto r_{\rm min}^{-3}\left(\frac{M_{\rm
1,max}}{M_{\rm 2,min}}\right)^{5/7},
\end{equation}
where $r_{\rm min}$ is the radius of influence of the central 
BH, while $M_{\rm
2,min}$ and $M_{\rm 1,max}=M_0$ are the minimum
and maximum BH masses from the distribution 
function. For a typical case where $M_2\ll M_1$, the 
merger rate of small BHs is lower than the rate of 
mergers with the central BH and tends to it in the 
limit $M_{\rm 1,max}\to M_0$. Thus, the mergers of relatively 
low-mass BHs (the mass boundary is defined by the 
condition $\rho_{\rm SN}(z,M)>5$) with the central BH near 
the cluster center $r\simeq r_{\rm min}$ make the largest contribution
to the gravitational radiation, while the rate of 
such mergers was calculated above. 

As has already been noted above, the present-day 
structured galaxies could include $\sim500$ BH clusters. 
These clusters sink to the galactic centers under 
dynamical friction and their central BHs merger together.
These mergers occur simultaneously with the 
BH mergers inside the clusters that we considered 
above. Distinctive signatures of the mergers of central 
BHs are their high masses, $M_0\simeq1.3\times10^5M_{\odot}$, and, 
accordingly, large amplitudes and low frequencies of 
the gravitational signals by which these mergers can 
be separated from the mergers of lower-mass BHs in 
observations. An accurate calculation of the rate of 
gravitational wave bursts requires cumbersome numerical
simulations of the galaxy merging process 
(allowance for the merger tree) and the secular BH 
settlement toward the galactic centers. Here, we give 
a simple estimate for the merger rate of central BHs 
with masses higher than $10^5M_{\odot}$ 
from the observable 
Universe: 
\begin{equation}
\frac{d^2N_{5}}{dtdz}\sim \frac{4\pi}{3}\frac{N}{t_0\Delta
z}(ct_0)^3~n_g \sim10\!
\left(\frac{n_g}{10^{-2}\mbox{Mpc$^{-3}$}}\right)
\!\left(\frac{t_0}{1.3\times10^{10}\mbox{yrs}}\right)^2\left(\frac{\Delta
z}{10}\right)^{-1} \!\left(\frac{N}{500}\right)\mbox{~yr$^{-1}$},
\nonumber
\end{equation}
where $n_g$ is the mean number density of structured 
galaxies $N\sim500$ is the mean number of mergers per 
galaxy, and $\Delta
z\sim10$ is the characteristic duration of 
the merger epoch in redshifts. Thus, the merger rate of 
massive central BHs with $M>10^5M_{\odot}$
is also fairly 
high and the gravitational wave bursts from these 
mergers can be detected by LISA. 

\newpage
 
\section*{DISCUSSION}

At present, there is no single, fully justified scenario
for the formation of supermassive BHs. Many 
different scenarios, both astrophysical and cosmological
ones, that could give rise to supermassive BHs 
have been developed and identifying the most likely 
mechanisms is of paramount importance. Future information
from the LISA gravitational wave detector 
will provide one of the possibilities for solving this 
problem, since existing models predict different rates 
and shapes of gravitational wave bursts. 

Our paper is based on the cosmological formation 
mechanism of the clusters of massive primordial BHs 
at a pregalactic expansion phase of the Universe developed
by Rubin et al. (2000, 2001), Khlopov and Rubin (2004), and Khlopov et al. (2005). We considered
two possibilities. First, gravitational waves 
are generated during the mergers of primordial BHs 
in clusters. Second, the BH clusters themselves are 
involved in the hierarchical clustering of dark matter, 
along with ordinary protogalaxies, being members of 
large galaxies. Under dynamical friction, the clusters 
of massive primordial BHs settle toward the galactic
centers, where they merge together to produce 
supermassive BHs, which is also accompanied by 
gravitational wave bursts. 

Based on this model, we found the rate of gravitational
wave bursts generated during the mergers
of primordial BHs in clusters. The hypothesis 
that this process is the main formation mechanism 
of supermassive BHs at the galactic centers allows 
the possible number of primordial BH clusters in a 
typical galaxy to be estimated. This hypothesis is of 
fundamental importance in calculating the possible 
rate of gravitational wave bursts from BH mergers 
detectable by LISA. The rate of gravitational wave 
bursts turned out to be high enough for their detection 
over the LISA operation time. 

Our main result is the conclusion that the redshift
distribution of bursts based on the cosmological
scenario differs significantly from the distribution 
given by the astrophysical models of late BH formation
in protogalaxies. Therefore, future observations 
of gravitational wave bursts will allow the dilemma of 
whether the massive BHs are galactic or cosmological
in origin to be solved in principle. In particular, 
for the cosmological origin, massive BHs could affect 
significantly the formation of large-scale structure. 

Note also that the formation mechanism of BH 
clusters suggested by Rubin et al. (2000, 2001), 
Khlopov and Rubin (2004), and Khlopov et al. (2005) 
can also be tested in principle based on ordinary 
optical observations. A number of BH clusters remained
outside galaxies should appear as dwarf 
galaxies (spheroids) with a sharp central density 
spike. Searching for such objects is of great interest
for this model. The dwarf spheroidal galaxies 
known to date have no central density spikes and, 
hence, could not be produced by this mechanism 
(Dokuchaev et al. 2008) but originated from ordinary 
density perturbations. It may also well be that BH 
clusters can be among the observed, but not yet 
identified powerful X-ray sources (Gao et al. 2004). 
The X-ray emission can be generated when baryons 
are accreted onto intermediate-mass BHs (Mapelli 
et al. 2008). 

\newpage

\section*{ACKNOWLEDGMENTS}

This work was supported by the Russian Foundation
for Basic Research (project nos. 06-02-16029 
and 06-02-16342) and by grants for support of 
leading scientific schools (4407.2006.2 and 5573.2006.2).

\section*{REFERENCES}

1. A. Cattaneo, J. Blaizot, J. Devriendt, and B. Guiderdoni,
Mon. Not. R. Astron. Soc. 364, 407 (2005). 

2. T. Damour and A. Gopakumar, Phys. Rev. B 73, 
124 006 (2006). 

3. V. Dokuchaev, Yu. Eroshenko, and S. Rubin, Grav. 
Cosmol. 99, 11 (2005). 

4. V. Dokuchaev, Yu. Eroshenko, and S. Rubin, 
arXiv:0709.0070v2 (2007). 

5. V. I. Dokuchaev, Yu. N. Eroshenko, and S. G. Rubin, 
Astron. Zh. (2008, in press); arXiv:0801.0885. 

6. L. S. Finn and K. S. Thorne, Phys. Rev. D 62, 124 021 
(2000). 

7. Y. Gao, Q. D. Wang, P. N. Appleton, and R. A. Lucas, 
Astrophys. J. 596, L171 (2003). 

8. L. P. Grishchuk, V. M. Lipunov, K. A. Postnov, et al., 
Usp. Fiz. Nauk 171, 3 (2001) [Phys. Usp. 44,1 
(2001)]. 

9. M. Yu. Khlopov and S. G. Rubin, Cosmological Pattern
 of Microphysics in the Inflationary Universe 
(Kluwer Acad. Publ., Dordrecht, 2004), Vol. 144. 

10. M. Yu. Khlopov, S. G. Rubin, and A. S. Sakharov, 
 Astrophys. Phys. 23, 265 (2005). 

11. LISA, http://lisa.jpl.nasa.gov.

12. M. Mapelli, B. Moore, L. Giordano, et al., Mon. Not. 
R. Astron. Soc. 383, 230 (2008).
 
13. T. Di Matteo, J. Colberg, V. Springel, et al., 
arXiv:0705.2269v1 (2007). 

14. H. Mouri and Y. Taniguchi, Astrophys. J. 566,L17 
(2002). 

15. S. G. Rubin, M. Y. Khlopov, and A. S. Sakharov, Grav. 
Cosmol. S 6, 51 (2000).
 
16. S. G. Rubin, A. S. Sakharov, and M. Y. Khlopov, 
JETP 92, 921 (2001).
 
17. A. Sesana, F. Haardt, P. Madau, and M. Volonteri, 
Astrophys. J. 623, 23 (2005).
 
18. L. Spitzer and W. C. Saslaw, Astrophys. J. 143, 400 
(1966). 

19. M. Volonteri, G. Lodato, and P. Natarajan, 
arXiv:0709.0529 (2007). 

20. C. M. Will, Astrophys. J. 611, 1080 (2004). 

\end{document}